# Novel Feature of Liquid Dynamics via Improvements in meV-Resolution Inelastic X-Ray Scattering

Alfred Q. R. BARON* & Daisuke ISHIKAWA

*Materials Dynamics Laboratory, RIKEN SPring-8 Center, 1-1-1 Kouto Sayo, Hyogo 679-5148 JAPAN*
*Precision Spectroscopy Division, CSRR, SPring-8/JASRI, 1-1-1 Kouto Sayo, Hyogo 679-5198 JAPAN*

We describe how improvements in methodology and instrumentation for meV-resolved inelastic x-ray scattering (IXS), coupled with a fresh examination of older theory, allow identification of interaction between the quasi-elastic and acoustic dynamical modes in liquid water. This helps explain a decades old controversy about the appearance of additional modes in water spectra, and provides a strong base from which to discuss new phenomena in liquids on the mesoscale.

## 1. Introduction

Liquid dynamics on the mesoscale - over ~nm correlation lengths - is where a transition from continuum behavior to atomistic behavior should occur. It is a fascinating subject, but also frustrating. Early work, through the 70's and 80's, was comprehensive and painted an excellent qualitative picture of the dynamics of liquids ([1-3] and references there-in), spurred, in part, by the invention of the laser and the light-scattering experiments that thereby became possible. However, as shorter wavelength probes began to be used - first neutrons, and then, later, x-rays - results from earlier work were discarded, or morphed into complex formalisms, possibly driven by a desire to offer something new. While new things do appear on short length scales, older models also make predictions for the mesoscale. It is then important to understand the range of validity of the previous work to be able to determine where new science may emerge. Particular subjects of interest for liquids, for example, include the often observed change in the dispersion of acoustic modes as their frequency exceeds the inverse relaxation time of the liquid, the huge increase in quasi-elastic scattering that is sometimes observed relative to measurements at long wavelengths, and the appearance, or not, of transverse or shear modes, that are forbidden in liquids on longer length scales, *etc*.

The mesoscale dynamics of liquid water is a case in point. Given the importance of water, it has been the subject of many investigations, including both inelastic neutron scattering (INS) and inelastic x-ray scattering (IXS) ([4-9] are a few references relevant to the present discussion - the field is large). While the data in similar conditions looks consistent, there is a surprising disparity of interpretation (see figure 1): some authors suggest the mesoscale data (momentum transfers of Q ~1 to 4 nm-1 corresponding to correlation lengths, $2\pi/Q$, from ~6 to 1.5 nm) supports the appearance of a propagating transverse mode [5], others suggest the presence of an anti-crossing similar to that seen in solids [6,8], and still other work (even by some of the same authors) suggests there is nothing anomalous [7]. These viewpoints have never been reconciled, but there has been a tendency to favor an interpretation in terms of the appearance of a propagating transverse mode[9]. This issue becomes especially important given the prominent position that the presence, or not, of propagating transverse dynamics can sometimes take in discussion of microscale fluid dynamics.

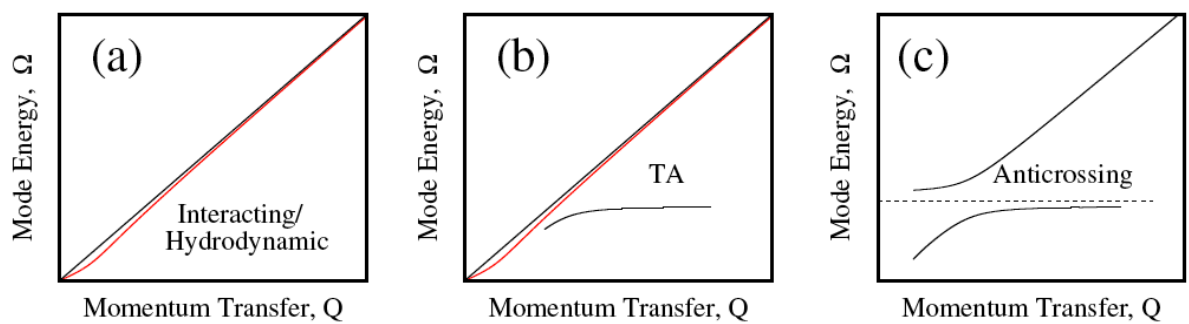

**Figure 1.** Schematic representation of several possible models for acoustic mode dispersion that may be indicated by choosing different approaches to fitting measured spectra (see text). A hydrodynamic model, (a), which includes interaction between quasi-elastic scattering and the acoustic mode, provides better fits to high-resolution spectra than models with an assumed transverse-acoustic (TA) mode (b) or an anti-crossing (c), even while it has the smallest number of degrees of freedom [15]. The red lines in (a) and (b) qualitatively show how the dispersion may be modified by viscoelasticity (see the discussion at the end of the article). For simplicity of display, the sound velocity used for the black lines is the fast or high-frequency sound velocity appropriate at larger momentum transfers.

---

* baron@spring8.or.jp

## 2. Improved Technique

At the RIKEN Quantum NanoDynamics Beamline, BL43LXU [10,11], see figure 2, we have been working to improve understanding of liquids on the mesoscale. This is experimentally non-trivial as it combines the need to measure at small angles (small momentum transfers) where background can be large, with a need for the best energy resolution as energy scales are small, with a relatively low count rate, as the dynamical signal typically scales as the square of the momentum transfer. Experimental and methodological improvements at BL43LXU include reducing the energy resolution to 0.8 meV [12] - the best now available for IXS - improving the momentum resolution [13], and improving the measurement of the full resolution function, including its tails [14], all in a stable and high-flux setup. The present discussion brings these together to demonstrate a feature of liquid dynamics, built into even elementary theory [3], that has not been previously recognized: in addition to a diffusive/relaxation component, and in addition to intrinsically broad acoustic modes, the dynamical response of a liquid fundamentally includes an interaction between the quasi-elastic component and the acoustic mode [15].

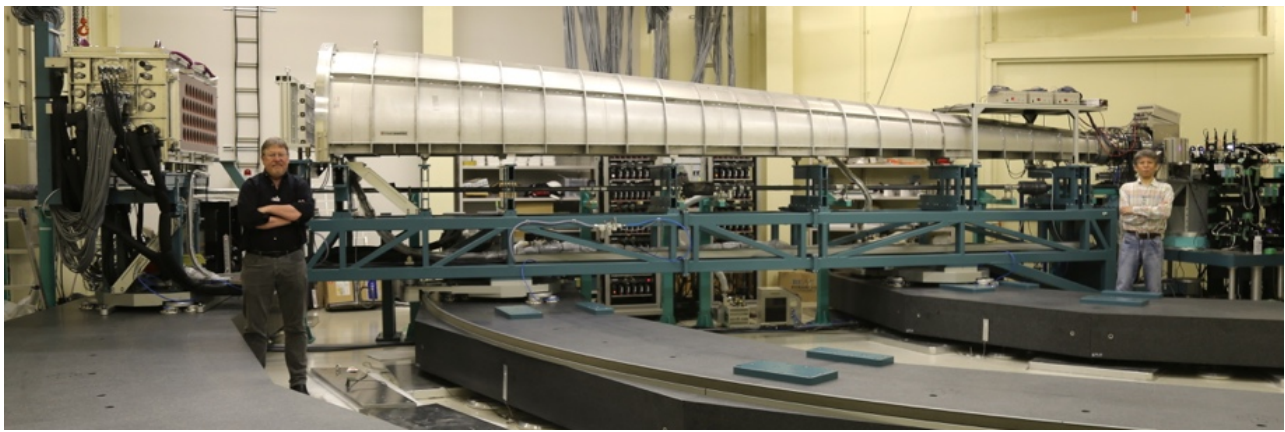

**Figure 2:** The high-resolution spectrometer at BL43LXU. Samples are placed on the stages at the far right and the meV analyzers are in the chamber at the far left. Two scientists (A. Baron, left, and D. Ishikawa, right) are shown for scale.

## 3. How are liquids different?

Liquids are one of several phases of matter which are often investigated, and it is useful to compare these phases from the viewpoint of their dynamical response. In (most!) crystalline solids, the atoms vibrate about fixed average positions and a simple harmonic, or ball and spring (also called a Born-von Karman) model, is often a shockingly good approximation. The level of complexity increases when one considers phonons interacting with other phonons or other (electronic, magnetic) systems, which may cause the phonons to develop complex line-shapes. While, rigorously speaking, such interaction requires solution of the full interacting Hamiltonian, in fact for interpretation of phonon spectra an approximation is often made: non-harmonic phonon lines are modelled with what is called a damped harmonic oscillator (DHO) line-shape, which is the lowest order approximation to the imaginary part of the relevant interacting Green's function with the correct symmetry, and allows for a finite linewidth [11,16].

The dynamics of glasses can be more complicated than that of crystals. Like crystals, atoms in a glass also vibrate about fixed positions (on the time scale of IXS measurements[†])However, the lack of long-range order in a glass means the mathematical technique, Bloch's theorem, used for periodic crystals no longer applies, and simple solutions to the dynamical problem in glasses are not available. However, practically, one can often approximate the dynamical response on the nm length scale as the sum of an elastic line from the static disorder and a broad acoustic mode modelled as a DHO. Thus, while there are many interesting questions related to, e.g., the momentum dependence of the linewidth, or the ratio of phonon scattering to elastic scattering (the "non-ergodicity" parameter) *etc*., a simple line shape, the sum of an elastic peak and a DHO acoustic mode, often provides a very good approximation to meV-scale spectra of glasses at small momentum transfers.

The atomic dynamics of liquids is different than glasses, as liquids are not only disordered, but that disorder changes on the time scale of the atomic vibrations: the time scale of atomic diffusion/structural relaxation is similar to that of acoustic mode vibrations. This is immediately clear in the finite quasi-elastic linewidth of liquid spectra on the meV scale, which is the most obvious feature distinguishing the dynamical spectral of liquids and glasses on the mesoscale. However, could there be something more?

---

[†] Atoms in glasses have long range motion on long (>~100s) time scales, but are effectively static over the ~ps time scales investigated by IXS.

## 4. Hydrodynamical Interaction

A starting point to understand the meso-scale dynamics of liquids is classical hydrodynamics. The Navier-Stokes equation that governs the macroscopic dynamics of liquids with well-defined thermodynamic parameters - thermal conductivity, thermal expansion, diffusivity, heat capacity, and viscosity - can be extended to short length scales by assuming small deviations from equilibrium conditions. This leads to a hydrodynamical solution for the dynamic structure factor, $S(Q,\omega)$ (the quantity measured in IXS and INS, where $Q$ is the momentum transfer and $\omega$ the energy transfer) which can be found in texts (e.g. [3,17]). This solution is a "Rayleigh-Brillouin triplet" with a quasi-elastic Rayleigh peak at zero energy transfer bracketed by the Stokes and anti-Stokes peaks of the dispersing longitudinal acoustic, Brillouin, mode, and is given by eqn. 1 (equivalent to eqn. 5.3.15 of [3])[‡].

$$S(Q,\omega) = \frac{S(Q)}{\pi}\left[ I_0 \frac{z_0}{\omega^2 + z_0^2} + I_1 \frac{2z_1\Omega^2}{(\omega^2 - \Omega^2)^2 + 4z_1^2\omega^2} + I_0 z_0 \frac{\Omega^2 - \omega^2}{(\omega^2 - \Omega^2)^2 + 4z_1^2\omega^2} \right] \qquad (1)$$

The quasi-elastic peak at zero energy transfer, the first term in the brackets in eqn. 1, is a Lorentzian, as one expects for simple diffusion or relaxation. However, the side peaks from the second and third terms are neither Lorentzian, nor DHOs: the second term in eqn. 1 is a conventional DHO but the third term creates an asymmetric response about the acoustic mode frequency. The physical origin of the asymmetry has not been clear, and, perhaps because of that, the asymmetry was dropped from the discussion of liquid water [4–6,8,9], and the default model for fitting spectra in the frequency domain became the L+DHO model (just the first two terms of eqn. 1), sometimes augmented by an additional DHO mode of arguable origin (e.g. [5,6,8,9]) leading to an L+2DHO model.

## 5. Results and Discussion

Motivated by a long-standing interest in the form of the response function for liquids, and by experimental improvements at BL43LXU, we returned to measure water with better energy and momentum resolution than previously used. The results were quite striking: we found that for all the spectra between 0.8 < Q < 4.2 nm$^{-1}$ the data were significantly better fit with the full hydrodynamic form of the response, eqn. 1, than with the L+DHO model. An example is shown in figure 3, where each term is plotted. In fact, even including a second DHO mode, an L+2DHO model (so allowing three additional free parameters relative to the L+DHO) the fits were, *at best*, comparable to the fits with eqn. 1, and sometimes significantly worse (see discussion in the supplemental materials for [15]). Considering Ockham's razor (choose the simplest explanation that fits the data), and support by theory, there is then excellent reason to use eqn. 1 to fit spectral data from ambient water on the mesoscale [15].

It then becomes a question how to interpret the third term. Here we are aided by several things. First, and most obviously, direct inspection of eqn. 1 shows the 3$^{rd}$ term responds to not only the parameters of the acoustic line, but also the parameters of the Lorentzian quasi-elastic line – thus it represents some interplay between the acoustic and quasi-elastic response. Second, the asymmetry is reminiscent of the Fano profile seen when a resonant response overlaps a continuum background – here the acoustic mode is the resonance and the tail of the quasi-elastic mode acts as the continuum. Finally, eqn. 1 can be converted to a form previously derived to explicitly describe coupling between different dynamical modes [18].

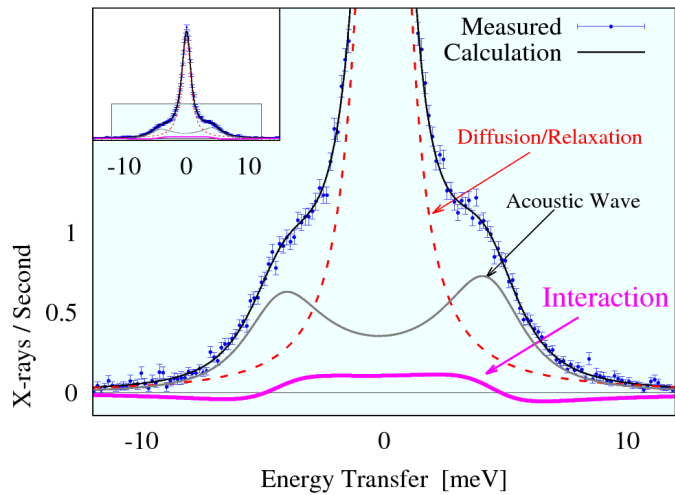

**Figure 3:** A spectrum from pure water at 301K at Q=2.5 nm$^{-1}$. The various contributions are as indicated. A fit with the full interacting model (black line) gives small residuals (reduced chi-squared 0.92, probability>0.5) while a fit without interaction (reduced chi-squared 1.56, probability<10$^{-5}$) give larger and highly correlated residuals. See text and [15]

---

[‡] The quantities in the equation are the widths of the quasi-elastic and acoustic modes, $z_0$ and $z_1$, the intensities of those modes, $I_0$ and $I_1$. The acoustic mode frequency is $\Omega$, so the sound speed is $\Omega/Q$, while $S(Q)$ is the energy integral of $S(Q,\omega)$, so that one requires $I_0 + I_1 = 1$.

Thus we have very good incentive to consider the additional term as being due to the interaction of the quasi-elastic and acoustic modes in liquids [15].

The use of the correct form for the response has significant implications for the interpretation of data. In particular, one of the interesting aspects of liquid dynamics is the phenomenon of "positive dispersion" or "fast sound" or "high-frequency sound". This refers to the fact that the acoustic mode energy in a liquid often increases (disperses) faster with increasing momentum transfer than expected from the macroscopic (ultrasonic) sound velocity. This is very different than what is usually observed in a solid, where the acoustic mode energy increases at or more slowly than the macroscopic sound velocity. Positive dispersion in a liquid is often discussed as being due to a solid-like response of the liquid appearing as the acoustic mode frequency exceeds the (inverse) relaxation time of the liquid: the liquid does not have an opportunity to relax on the time scale of the acoustic vibration so the material behaves more stiffly. This can be made more formally exact in a viscoelastic model of liquid response (e.g. [2,3])

Figure 4 shows the speed of sound determined when either the full interacting model of eqn. 1 is used, or the, rather more poorly fitting, L+DHO form. The red straight line indicates the macroscopic sound velocity, 1.5 km/s, for ambient water, while the saturated fast sound velocity in is given by the other lines. The difference in the two models is significant, and in fact becomes fractionally quite large as one moves into the transition region, with the difference in positive dispersion (the sound velocity increase relative to the macroscopic value) being nearly a factor of two at ~1 nm$^{-1}$. One also notes that the saturation of the sound velocity occurs at different momentum transfers for the two models. This indicates the importance of using a correct model when discussing liquid properties.

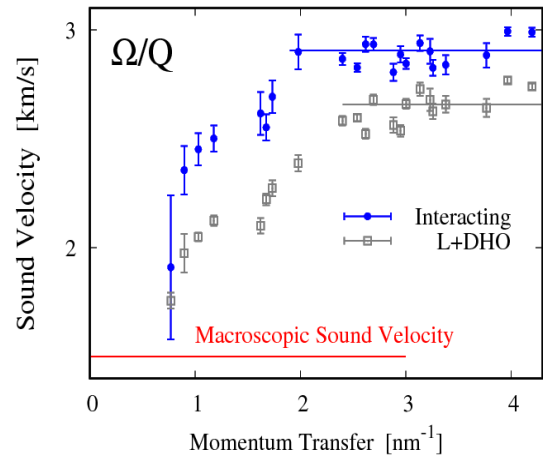

**Figure 4:** Sound velocity as determined using either the interacting model or a L+DHO model. The differences are significant. See text for discussion

Finally, we discuss the extremely large intensity of the quasi-elastic peak in the spectra (see the inset of figure 3), as this is perhaps the most surprising aspect of the water spectra from the viewpoint of results using longer wavelength probes. In light scattering at small momentum transfers (Q~0.02 nm$^{-1}$), the quasi-elastic peak is only about 3% of the phonon peak intensity [19], in agreement with a well-known relationship that the intensity ratio $I_0/I_1$, called the Landau-Placzek ratio, is given by the difference of the ratio of specific heat capacities from unity as $I_0/I_1 = C_p/C_v - 1$. Meanwhile, the quasi-elastic peak in figure 3 is nearly 300% of the acoustic mode intensity for IXS - a dramatic change. Returning to older theoretical work (e.g. [2]) one finds that, when the fast sound velocity is large, and the heat capacity ratio is near unity, as it is for ambient water, the quasi-elastic intensity is dominated by the relaxation causing the fast sound. In fact, a relation for the relative intensity of the quasi-elastic peak in terms of the magnitude of the fast sound [2] is in good agreement with the observed fast sound [15], suggesting that indeed the increased quasi-elastic scattering is due to the structural relaxation causing the fast sound.

## 6. Conclusion

In sum, with the improved capabilities of BL43LXU, we have gone back and re-investigated the properties of ambient water. Different than previous work, we find that it is very important to consider the full (to second order) response predicted by hydrodynamics, *including* a term related to the interaction of the acoustic and quasi-elastic scattering. This suggests the previous treatments suggesting additional modes may be present corresponding to either a propagating transverse mode or an anti-crossing, might be reconsidered. Further, we find good reason to associate the huge quasi-elastic peak with exactly the relaxation process that leads to fast sound. Taken together, the experimental improvements coupled with careful modelling are then expected to provide a good basis for additional analysis. This is an important and necessary step toward clearly identifying where new dynamical properties of liquids may emerge.

### Acknowledgements

AB is grateful to Masanori Inui of Hiroshima University for constructive comments on an earlier version of this paper. The meV IXS measurements were made at the RIKEN Quantum NanoDynamics Beamline, BL43LXU, of the RIKEN SPring-8 Center.

### References

[1] L. D. Landau and E. M. Lifshitz, *Fluid Mechanics*, Second Edition (Elsevier, Oxford, 1959).

[2] R. D. Mountain, J. Res. Natl. Bur. Stand. -A. Phys. Chem. **70A**, 207 (1966).
   https://pubmed.ncbi.nlm.nih.gov/31823990/

[3] J. P. Boon and S. Yip, *Molecular Hydrodynamics* (Dover Publications, Mineola, New York, 1980).

[4] J. Teixeira, M. C. Bellissent-Funel, S. H. Chen, and B. Dorner, Phys. Rev. Lett. **54**, 2681 (1985). https://link.aps.org/doi/10.1103/PhysRevLett.54.2681

[5] F. Sette, G. Ruocco, M. Kirsch, C. Mascivecchio, R. Verbini, and U. Bergmann, Phys. Rev. Lett. **77**, 83 (1996). https://link.aps.org/doi/10.1103/PhysRevLett.78.976

[6] C. Petrillo, F. Sacchetti, B. Dorner, and J.-B. Suck, Phys. Rev. E **62**, 3611 (2000). https://link.aps.org/doi/10.1103/PhysRevE.62.3611

[7] G. Monaco, A. Cunsolo, G. Ruocco, and F. Sette, Phys. Rev. E **60**, 5505 (1999). https://link.aps.org/doi/10.1103/PhysRevE.60.5505

[8] F. Sacchetti, J.-B. Suck, C. Petrillo, and B. Dorner, Phys. Rev. E **69**, 61203 (2004). https://link.aps.org/doi/10.1103/PhysRevE.69.061203

[9] A. Cunsolo, C. N. Kodituwakku, F. Bencivenga, M. Frontzek, B. M. Leu, and A. H. Said, Phys. Rev. B **85**, 174305 (2012). https://link.aps.org/doi/10.1103/PhysRevB.85.174305

[10] A.Q.R. Baron, SPring-8 Inf. Newsl. **15**, 14 (2010). http://user.spring8.or.jp/sp8info/?p=3138

[11] A. Q. R. Baron, in *Synch. Light Srcs. & FELS*, edited by E. Jaeschke, *et al.* (Springer International Publishing, Cham, 2016), pp. 1643-1757 See also arXiv 1504.01098. https://arxiv.org/abs/1504.01098

[12] D. Ishikawa, D. S. Ellis, H. Uchiyama, and A. Q. R. Baron, J. Synch. Rad. **22**, 3 (2015). https://doi.org/10.1107/S1600577514021006

[13] A. Q. R. Baron, D. Ishikawa, H. Fukui, and Y. Nakajima, AIP Conf. Proc. **2054**, 20002 (2019). https://aip.scitation.org/doi/abs/10.1063/1.5084562

[14] D. Ishikawa and A. Q. R. Baron, J. Synch. Rad. **28**, 804 (2021). https://doi.org/10.1107/S1600577521003234

[15] D. Ishikawa and A. Q. R. Baron, J. Phys. Soc. Japan **90**, 83602 (2021). https://doi.org/10.7566/JPSJ.90.083602

[16] B. Fak and B. Dorner, ILL Rep. 92FA008S (1992).

[17] N. H. March and M. P. Tosi, *Atomic Dynamics in Liquids* (Dover Publications, New York, 1976).

[18] K. H. Michel and J. Naudts, J. Chem. Phys. **68**, 216 (1978). https://aip.scitation.org/doi/abs/10.1063/1.435485

[19] C. L. O'Connor and J. P. Schlupf, J. Chem. Phys. **47**, 31 (1967). https://doi.org/10.1063/1.1711865